\documentclass[sigconf]{acmart}
\usepackage{bm}
\usepackage[ruled,linesnumbered]{algorithm2e}

\AtBeginDocument{%
  \providecommand\BibTeX{{%
    \normalfont B\kern-0.5em{\scshape i\kern-0.25em b}\kern-0.8em\TeX}}}

\setcopyright{acmcopyright}
\copyrightyear{2022}
\acmYear{2022}
\acmDOI{1.1111/1122333.1122333}

\acmConference[KDD 'XX]{KDD 'XX: ACM SIGKDD Conference on Knowledge Discovery and Data Mining}{XXXX, 20XX}{XXXX, XXX}
\acmBooktitle{KDD 'XX: ACM SIGKDD Conference on Knowledge Discovery and Data Mining,
  XXXXXXX, 20XX, XXXXXXXXX}
\acmPrice{15.00}
\acmISBN{978-1-4503-XXXX-X/18/06}

\usepackage{amsmath,amssymb}

\usepackage{algorithm}
\usepackage{algpseudocode}
\usepackage{multicol}
\usepackage{multirow}

\begin{document}
\fancyhead{}
\title{On Ranking Consistency of Pre-ranking Stage}

%
\author{Siyu Gu, Xiangrong Sheng}
\email{gusuperstar@gmail.com}
\affiliation{%
  \city{Beijing, China}
}
\renewcommand{\shortauthors}{Gu and Sheng, et al.}




 \begin{abstract}

Industrial ranking systems, such as advertising systems, rank items by aggregating multiple objectives into one final objective to satisfy user demand and commercial intent. Cascade architecture, composed of retrieval, pre-ranking, and ranking stages, is usually adopted to reduce the computational cost. Each stage may employ various models for different objectives and calculate the final objective by aggregating these models' outputs. 
The multi-stage ranking strategy causes a new problem - the ranked lists of the ranking stage and previous stages may be inconsistent. 
For example, items that should be ranked at the top of the ranking stage may be ranked at the bottom of previous stages. In this paper, we focus on the \textbf{ranking consistency} between the pre-ranking and ranking stages. Specifically, we formally define the problem of ranking consistency and propose the Ranking Consistency Score (RCS) metric for evaluation. We demonstrate that ranking consistency has a direct impact on online performance. Compared with the traditional evaluation manner that mainly focuses on the individual ranking quality of every objective, RCS considers the ranking consistency of the fused final objective, which is more proper for evaluation. Finally, to improve the ranking consistency, we propose several methods from the perspective of sample selection and learning algorithms. Experimental results on one of the biggest industrial E-commerce platforms in China validate the efficacy of the proposed metrics and methods. 

 \end{abstract}






\maketitle

\section{Introduction}\label{sec:intro}
\begin{figure*}[!t]
    \centering
    \includegraphics[width=0.8\textwidth]{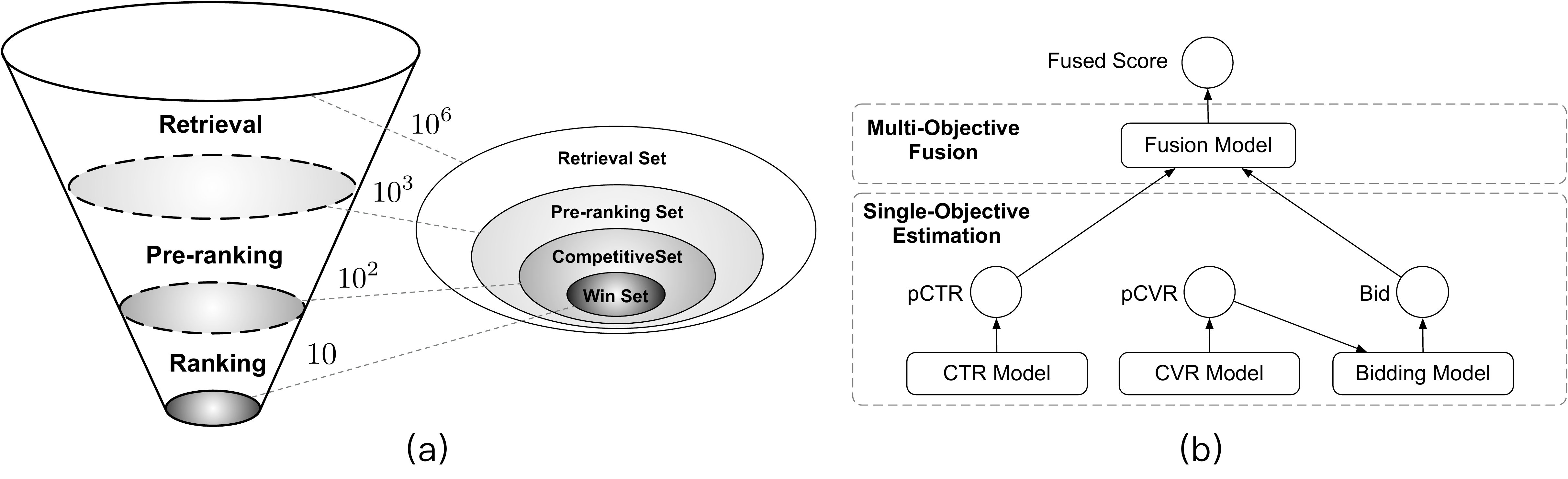}
    \caption{(a) An example of the cascade architecture, which consists of retrieval, pre-ranking and ranking stages. Each stage progressively refines the results from the preceding stage.
    (b) A multi-objective ranking system (advertising system), which fuses different objectives into the final score.}
    \label{fig:multisys}
\end{figure*}

Large-scale industrial ranking systems, such as advertising and recommendation systems, need to select a few items from a large corpus, usually consisting of billions of candidates~\cite{Pike-Burke0SG2018DelayedAggregatedAnonymousFeedback,CovingtonAS2016YouTubeDNN,ZhuJTPZLG2017OCPC}. 
Such systems have two main characteristics:
\begin{itemize}
    \item \textbf{Cascade architecture:} The system usually adopt the cascading architecture, composed of retrieval, pre-ranking, and ranking stages. As shown in Figure \ref{fig:multisys}(a), each stage takes a different magnitude of candidates, ranks these candidates, and then returns the top candidates for the next stage. In the display advertising system, the pre-ranking stage receives thousands of candidates and returns the best hundreds. In the ranking stage, complex models~\cite{zhou2019dien,JiaqiMa2018ModelingTR} are employed while simpler models~\cite{CovingtonAS2016YouTubeDNN,WangZJZZGCold} are adopted at previous stages. The cascade architecture allows the system to rank items from a vast corpus while still meeting the strict constraint of latency and computational cost.
    \item \textbf{Multiple objectives:} To satisfy users' diverse demand and the commercial intent, industrial system need to consider multiple objectives, such as eCPM (effective Cost Per Mille), CTR, GMV (Gross Merchandise Volume). For each stage, different objectives are usually estimated separately~\cite{zhou2019dien,GuSFZZ2021Defer} and then be fused to obtain the final score of the stage, as shown in Figure \ref{fig:multisys}(b).

\end{itemize}

Since the ranking stage adopts more complicated models than the pre-ranking stage, its ranking quality is better than the pre-ranking stage. Thus given the same candidate set, the ranked list provided by the ranking and pre-ranking stages may be different. We refer to the consistency of the ranking quality as the \textit{ranking consistency}. To show why the inconsistency is likely to happen in a multi-objective system, we give an toy example of an advertising system, as shown in Table~\ref{tab:rcexample}. Here bid and pCTR are two single objectives. The final fused objective is eCPM = bid * pCTR. We can see that although the consistency of all single objectives are kept (the order of pCTR/bid is consistent), the ranking inconsistency may still happen after fusing different objectives (the oder of eCPM is inconsistent).  

Note that the inconsistency can negatively impact the system. For example, items with high scores at the ranking stage but low scores at the pre-ranking stage are less competitive and will not be selected for the ranking stage, hurting the system's effectiveness. On the contrary, items with low scores at the ranking stage but high scores at the pre-ranking stage will not be selected for impression. In this scenario,  the computational resource of the ranking stage is wasted for inferior items, undermining efficiency. 

\begin{table}[t]
    \caption{A toy example of an advertising system, where eCPM = bid*pCTR is the final objective. Although all single objectives are consistent (the order of pCTR is consistent), the ranking inconsistency still happens after fusing pCTR and bid (the oder of eCPM is inconsistent).}
    
    \label{tab:rcexample}
    \resizebox{\columnwidth}{!}{{\begin{tabular}{c|cccc|cccc}
    \toprule
      {\multirow{2}{*}{sample id}} &\multicolumn{4}{c|}{Pre-ranking} & \multicolumn{4}{c}{Ranking} \\
      \cline{2-9}
      & bid    & pCTR & fused score & order & bid & pCTR & fused score & order  \\
    \midrule
    1    & 8 & 0.4 & 3.2 & 1& 8 &0.2& 1.6 & 3 \\
    2   & 6 & 0.5 & 3.0 & 2& 6 &0.5 &3.0 & 2 \\
    3     & 4 & 0.6 & 2.4 &3 & 4 &0.8& 3.2 & 1 \\
    \bottomrule
    \end{tabular}}}
 \end{table}
 
Ideally, if we use the ranking models to serve for both the pre-ranking and ranking stages, these two stages are consistent. Note that we can also define the \textit{Ranking Consistency Score} (RCS) in a similar way to measure the ranking consistency: we first use the ranking models to select the ideal win set directly from the pre-ranking set (simulate the process that ranking models serve for both the pre-ranking and ranking stage). Then we define the RCS as the average hit rate of the competitive set (selected by the pre-ranking models) and ideal win set. The computation of RCS is illustrated in Figure~\ref{fig:rcs_computation}.

One difficulty of computing RCS is that the production system can not log the predictions of ranking models on the pre-ranking set due to the constraint of latency. Thus we propose to use
an \textit{online simulator}, on which the ranking models are deployed. The online simulator does not serve for the main traffic thus has no latency constraint. For each request, it will be send to both production system and the online simulator and the online simulator logs the prediction score on the pre-ranking
set.

The proposed RCS can be used to evaluate the performance of pre-ranking models and help find models that causes inconsistency. To understand what factors that influence the ranking consistency, we conduct extensive experiments on our production environment with RCS as the metric.
We discover that both the \textbf{ranking quality} and \textbf{proxy-calibration} (the output of the pre-ranking model aligns with the ranking model) contributes to the ranking consistency. Thus, unlike the common practice that only consider the ranking quality (e.g., the
ranking quality of the CTR model), when choosing the best model of pre-ranking stage, both the ranking quality and proxy-calibration need to be considered.

Besides, we discover that the inconsistency is more severe on unexposed samples. 
To mitigate the inconsistency of unexposed samples, we aim to improve the pre-ranking stage from the perspective of sample selection and learning algorithms. For the training sample,
we propose to utilize samples in the pre-ranking set to train the pre-ranking models. For the learning algorithms, we propose to two methods: 1) distilling
the prediction scores on the pre-ranking set of ranking models and 2) directly learning the rank order of the ranking stage. The proposed method effectively improves the consistency and thus improve the performance of online system.

The contributions of this paper are summarized as follow:
\begin{itemize}
    \item We discover that the ranking consistency of pre-ranking and ranking stages has a strong impact on the performance of industrial systems. To measure the ranking consistency, we leverage recall metric and call it as Ranking Consistency Score (RCS). Different other works, we point out that RCS is not only better for evaluating pre-ranking models, but also can be used to reveal the inconsistent module of pre-ranking stage.
    \item To understand what factors that influence the ranking consistency, we conduct extensive experiments on our production environment with RCS as the metric. We discover that both the ranking quality and proxy-calibration of single objective contributes to the ranking consistency.
    \item To enable the pre-ranking stage to align with the ranking stage, we propose sample selection and learning algorithms. Experiment results show that the proposed methodology can greatly improve the ranking consistency. Up to now, the proposed methodology has been deployed on the display advertising systems of one of the biggest E-commerce platform in China. The online A/B testing validates the efficacy of the proposed methods.
\end{itemize}

\section{Preliminary}
In this section, we give a brief introduction about the multi-objective ranking system with cascade architecture. Then we discuss the issues of the common evaluation manner for pre-ranking models. 
\subsection{Multi-Objective Ranking System with Cascade Architecture}
Industrial system, such as search engines, recommendation systems, and advertising systems, usually adopts a cascade architecture consisting retrieval, pre-ranking and ranking stages, where each stage progressively refines the results from the preceding stage~\cite{CovingtonAS2016YouTubeDNN,HuangSSXZPPOY2020FacebookEBR,WangZJZZGCold}.  
\begin{itemize}
    \item The goal of the \textbf{retrieval stage} is to retrieve a small subset of candidate items from the large corpus called \textit{retrieval set}, usually with billions or trillions of items. Nearest neighbour search\cite{CovingtonAS2016YouTubeDNN} or Tree methods\cite{zhu2018tdm} are usually used.
    \item The input of the \textbf{pre-ranking stage} is the output obtained from the retrieval stage, called \textit{pre-ranking set}. The output of the pre-ranking stage is the \textit{competitive set}. A DNN model could be used in the pre-ranking stage, utilizing a small set of features and a condensed model architecture~\cite{WangZJZZGCold}.
    \item Different from the retrieval and the pre-ranking stage which usually takes billions or thousands of items into consideration, the \textbf{ranking stage} only needs to process hundreds of items. Therefore, the ranking models usually contain a large set of features describing the user, items, and context and use complicated architectures~\cite{zhou2018din,ShengZZDDLYLZDZ2021STAR,DBLP:conf/wsdm/BianWRPZXSZCMLX2022CAN}. The ranking stage gives the predictions of different objectives by different ranking models.
\end{itemize}

Given the prediction scores of different objectives, the system will fuse several single objectives into the final objective. For example,  advertising system rank items by the product of pCTR and bid, i.e., eCPM, and return the top $k$, e.g., $k=5$, items with the highest fused ranking score. The top $k$ items is referred to as \textit{win set}, which will be exposed to the users.

Our main goal is to study the ranking consistency between pre-ranking and ranking stage.  The pre-ranking set contains items ${D=\{d_1, \dots, d_n\}}$ with corresponding features ${X=\{x_1, \dots, x_n\}}$. $f_i$ denotes the learned pre-ranking model of the $i$-th objective and $f_i(x_j)$ denotes the prediction of the $j$-th item. $F_i(X) = \{f_i(x_1), \dots, f_i(x_n)\}$ is the score list for objective $i$ regarding the set $X$. After obtaining the predictions of all $M$ objectives $F_1(X), \dots, F_M(X)$, the system fuses these score into the final ranking score:
\begin{equation}
G(X) = g(F_1(X), \dots, F_M(X))
\end{equation}
where $g$ is the pre-defined fusion function. For example, in online advertising, $f_1$ is the pCTR model and $f_2$ is the bid, $g(F_1(X), F_2(X)) = F_1(X)*F_2(X)$ is the element-wise product.

\begin{figure}[!t]
    \centering
    \includegraphics[width=0.7\columnwidth]{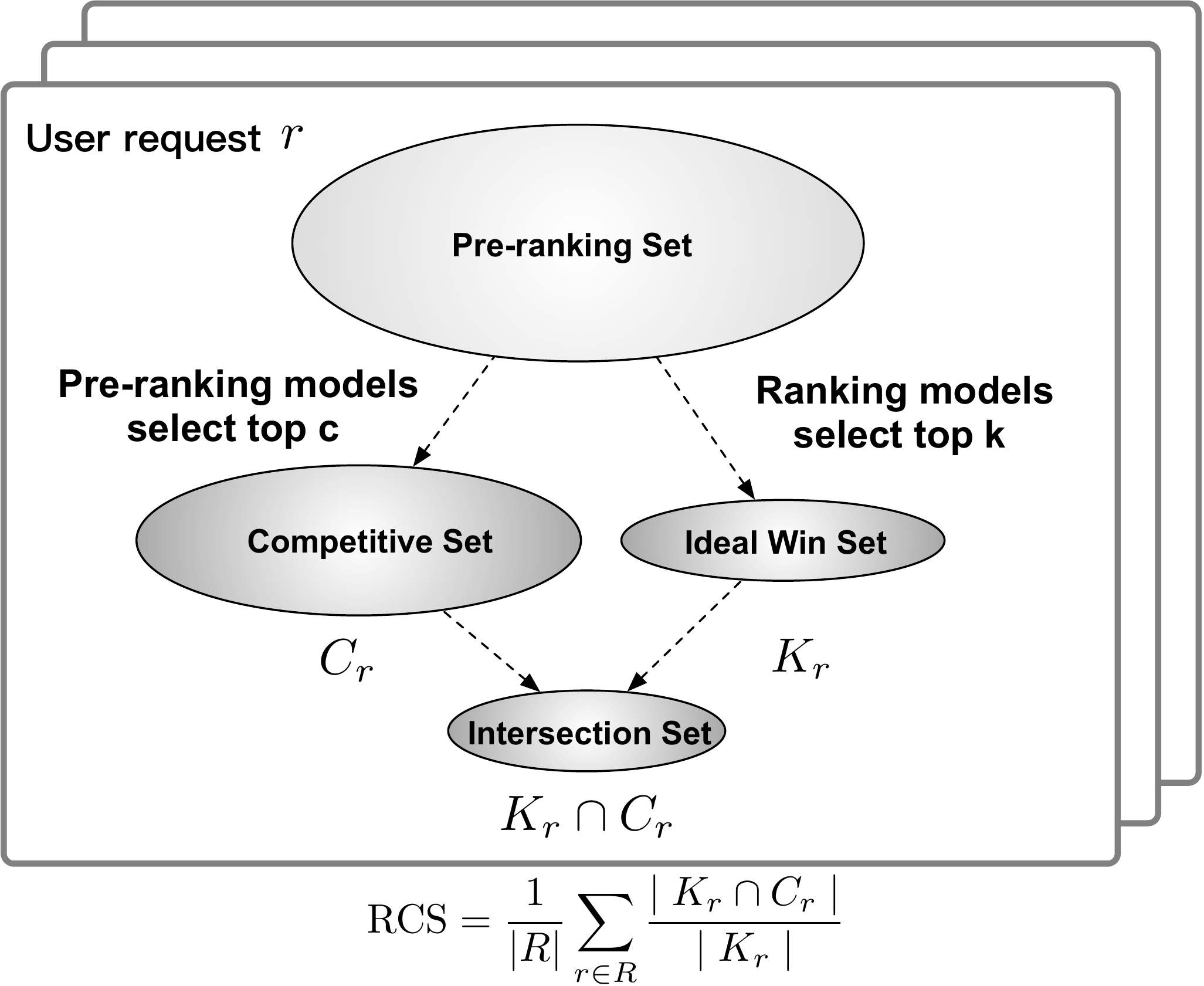}
    \caption{The computation of Ranking Consistency Score (RCS).}
    \label{fig:rcs_computation}
\end{figure}
\subsection{Issues of the Common Evaluation Manner for Pre-ranking Models}
The common offline evaluation manner is to measure the ranking quality of each single pre-ranking model (pCTR or any other model) using metrics like AUC or group AUC (GAUC)~\cite{ZhuJTPZLG2017OCPC,zhou2018din}. After that, one can choose the model with best ranking quality for online experiments. However, this offline evaluation manner evaluates different objective model individually, which does not consider the ranking consistency (after fusing all objectives) between pre-ranking and ranking stages. Although the ranking quality of each single model may be good, after fusing the output of different models, the ranking consistency may become worse, like the example shown in Table~\ref{tab:rcexample}. 


\section{Rethinking Ranking Consistency}

In this section, we propose to utilize the recall metric to measure the ranking consistency between the pre-ranking and ranking stages. We also present the implementation details it.

\subsection{The Ranking Consistency Score}

Instead of separately evaluating the ranking quality of each model, we should evaluate the ranking consistency (after fusing all objectives) between pre-ranking and ranking stages. More formally, we first define \textit{perfect consistency} as
\begin{equation}
\begin{aligned} 
    f_i(x) = \overline{f}_i(x), \forall x \in \mathcal{X}, \forall 1\le i\le M
\label{eq:perfect-consistent}
\end{aligned}
\end{equation}
where $\mathcal{X}$ represents all samples in the pre-ranking stages. $f_i$ and $\overline{f}_i$ are the pre-ranking and ranking models of the $i$-th objective, respectively. Equation~\ref{eq:perfect-consistent} means that pre-ranking and ranking models have the same prediction on all samples.
However, in real ranking systems, it is hard to achieve perfect consistency due to the different model capacity of these two stages. Additionally, 
perfect consistency is not necessary for ranking consistency. Ideally, if we use the ranking models to serve for both the pre-ranking and ranking stages, i.e., combining the two stages into one stage, then the system is consistent. Here, the win set selected by ranking models from the pre-ranking set is referred to as the \textbf{ideal win set}.
Since the role of the pre-ranking stage is to select the \textbf{competitive set} from the \textbf{pre-ranking set}, if we can guarantee the competitive set contains all items in the ideal win set, i.e., if Equation~\ref{eq:ranking-consistent} is ensured, 
\begin{equation}
\begin{aligned} 
    K_r \subseteq C_r, \forall r \in R
\label{eq:ranking-consistent}
\end{aligned}
\end{equation}
then the two stages are consistent. 
Here, $K_r, C_r$ are the ideal win set and competitive set for request $r$, respectively. $R$ represents all user requests.

Motivated by Equation~\ref{eq:ranking-consistent}, we propose to utilize Ranking Consistency Score (RCS) as the metric for evaluating the ranking consistency 
\begin{equation}
    {\rm RCS} = \frac{1}{|R|}\sum_{r\in R} \frac{\mid K_r \cap C_r \mid}{\mid K_r \mid},
    \label{eq:RCS}
\end{equation}
which is average ratio that items in ideal win set $K_r$ can be covered by competitive set $C_r$. It is a recall metric mentioned in ~\cite{XuMa2021TowardsAB,WangZJZZGCold}. Base on them, we clarify the calculation details and then point out that RCS is not only for evaluating pre-ranking models, but also
can be used to reveal the inconsistent module of pre-ranking stage.
In detail, the computation of Ranking Consistency Score (RCS) contains three steps. 
\begin{enumerate}
    \item Let $X_r$ denotes the pre-ranking set of request $r$, we can use ranking models to compute the fused ranking score on the pre-ranking set, denoted as $\overline{G}(X_r) = g(\overline{F}_1(X_r), \dots, \overline{F}_M(X_r))$. Then we select top $k$ items with highest score as the \textbf{ideal win set} $K_r$.
    \item Then we select the top $c$ items with highest fused score $G(X)$ as the \textbf{competitive set} $C_r$ for each request $r\in R$.
    \item Finally we compute the average ratio that items in ideal win set $K_r$ can be covered by competitive set $C_r$ as Equation~\ref{eq:RCS}.
\end{enumerate}

RCS can be seen as the recall metric on the pre-ranking set. Intuitively, RCS measures how well the pre-ranking model can select valuable items for the ranking stage, which matches the goal of the pre-ranking stage. Figure~\ref{fig:rcs_computation} illustrates the computation process of RCS. Note that the computation of  RCS depends on the parameter $c$ and $k$, where $c$ is the size of the competitive set and and $k$ is size of the win set. If the competitive set gets bigger (the ranking stage predicts more items), the RCS will also increase.

\begin{figure}[!t]
    \centering
    \includegraphics[width=\columnwidth]{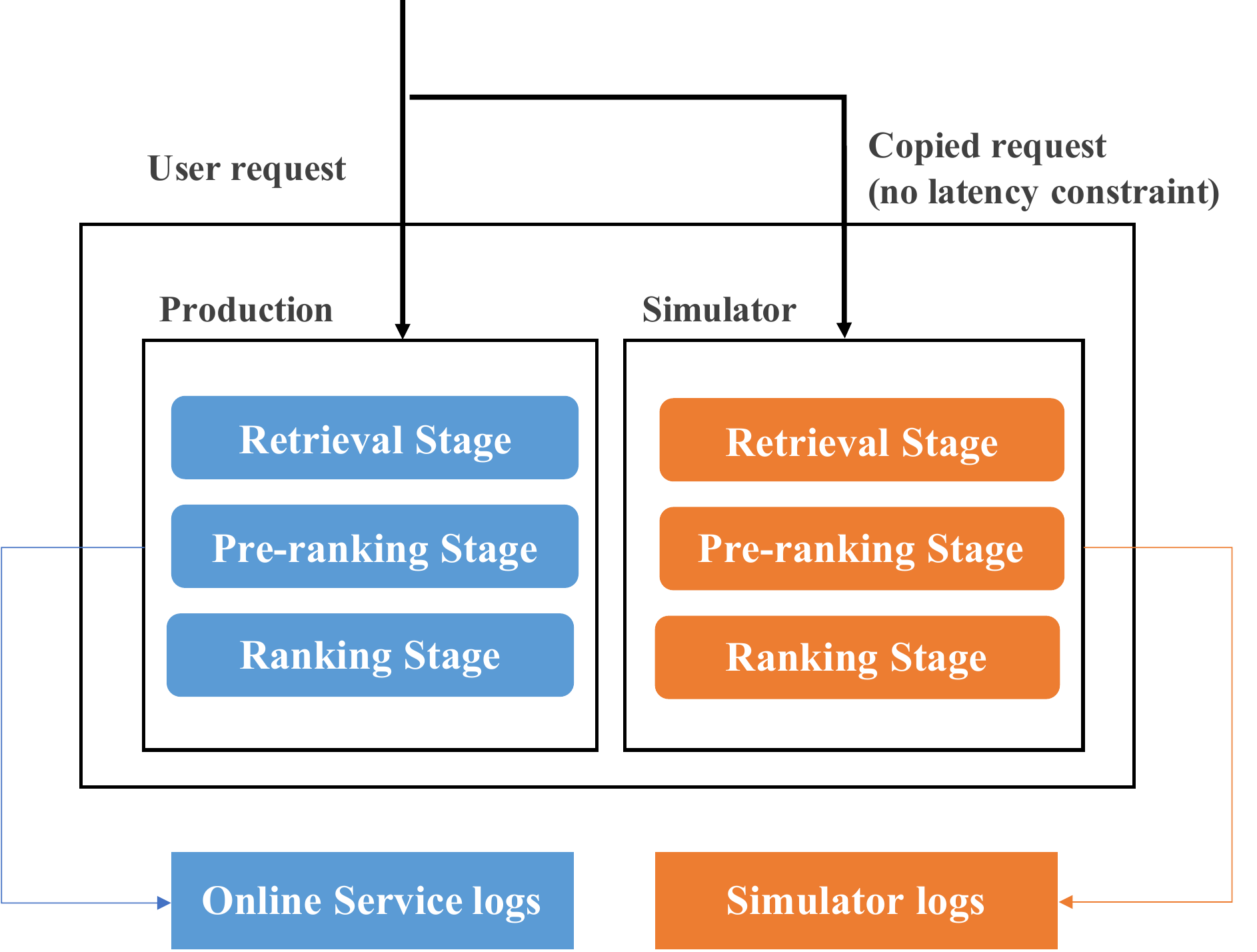}
    \caption{The online simulator that logs the predictions of ranking model on pre-ranking set.}
    \label{fig:simulator}
\end{figure}

To compute RCS, we need to get the fused score of ranking stages on pre-ranking set. Commonly, the industrial system only logs the predicted scores like pCTR on the competitive set. 
To obtain the fused score on pre-ranking set, we resend each request to an \textit{online simulator} to logs the prediction score on the pre-ranking set, as shown in Figure~\ref{fig:simulator}. The simulator is decoupled with the production system and have no latency constraint.
Using the online simulator, we can collect predicted scores on pre-ranking set for analysis. The simulator also enable us to monitor the ranking consistency of the online ranking system by computing RCS in real-time. The pseudocode of the RCS computation is shown in Algorithm~\ref{alg:rcs}. 

{\footnotesize
\begin{algorithm}[!t]

\caption{A SQL-style Pseudocode of RCS metric.}
\label{alg:rcs}

\begin{algorithmic}

\State \textbf{INPUT: } $k, c$, SIMULATOR\_LOG, ONLINE\_SERVICE\_LOG
\State \textbf{OUTPUT: } RCS 


\State \textbf{SELECT} SUM(IF(C.pv IS NOT NULL, $1$, $0$)) / SUM(K.pv) \textbf{AS} RCS

\State \textbf{FROM} (
\State \quad \quad \textbf{SELECT * FROM} (
\State \quad \quad \quad \quad \textbf{SELECT} request\_id, pv, item\_id,
\State \quad \quad \quad \quad \quad \quad \textbf{ROW\_NUMBER() OVER (PARTITION BY} request\_id \State \quad \quad \quad \quad \quad \quad \textbf{ORDER BY} g\_score desc) \textbf{AS} rank\_pos 
\State \quad \quad \quad \quad \textbf{FROM} SIMULATOR\_LOG
\State \quad \quad )
\State \quad \quad \textbf{WHERE} rank\_pos <= $k$
\State )$K$ {\color{teal}\,\# ideal win set}
\State \textbf{LEFT JOIN}(
\State \quad \quad \textbf{SELECT * FROM} (
\State \quad \quad \quad \quad \textbf{SELECT} request\_id, pv, item\_id,
\State \quad \quad \quad \quad \quad \quad \textbf{ROW\_NUMBER() OVER (PARTITION BY} request\_id \State \quad \quad \quad \quad \quad \quad \textbf{ORDER BY} g'\_score desc) \textbf{AS} pre\_rank\_pos
\State \quad \quad \quad \quad \textbf{FROM} ONLINE\_SERVICE\_LOG
\State \quad \quad )
\State \quad \quad \textbf{WHERE} pre\_rank\_pos <= c
\State )C {\color{teal}\,\# competitive set}
\State \textbf{ON} K.request\_id = C.request\_id \textbf{AND} K.item\_id = C.item\_id

\end{algorithmic}

\end{algorithm}
}

\subsection{Evaluating Single-objective Models}
 
The common industrial practice for optimizing the pre-ranking stage is to iterate models of different objectives, e.g., CTR and CVR, separately. However, since the outputs of different models are fused to obtain the final ranking score, each model interacts with the others. One way to measure the goodness of a single-objective model is to fix other models and modify a single-objective model to see the improvements on RCS. 
By using RCS as the metric, we identify factors that are helpful for improving the ranking consistency when iterating the single-objective model.
In detail, the \textbf{ranking quality} and \textbf{proxy-calibration} (the output of the pre-ranking model aligns with the ranking model) 
are closely related to the ranking consistency. 

\subsubsection{Proxy-Calibration.}
Since the prediction score of each single objective is used to compute the final fused score, the scale of the predictions should also be aligned, otherwise it may also cause inconsistency. Table~\ref{tab:rcexample} gives an example, although the ranking quality of the pre-ranking model are aligned, the rank order changes after multiplying the bid due to the misalignment of the absolute pCTR value. For simplicity, we assume that all models are probabilistic model that predict a probability estimate for each objective.
Motivated by Equation~\ref{eq:perfect-consistent}, we define the \textit{proxy-calibration} as the problem of prediction of pre-ranking model matches the prediction of ranking model.

Note that the concept of the proxy-calibration is similar to \textit{calibration}~\cite{GuoPSW2017OnCalibration} that measures the alignment of the predicted score with the true probability. However, calibration relies on the user response label, e.g., the click, and it is impossible to obtain the label for unexposed samples (most samples of the pre-ranking set is unexposed). Proxy-calibration can be seen as the extension of the calibration concept on the pre-ranking set with the output of ranking model as the ground truth. Thus, the common metrics used to assess the calibration performance, e.g., like Expected Calibration Error (ECE)~\cite{GuoPSW2017OnCalibration} and Predicted CTR Over CTR (PCOC)~\cite{ShengZZDDLYLZDZ2021STAR}, can also be used to measure the proxy-calibration performance.

\subsubsection{Ranking Quality.} 
Proxy-calibration measures the alignment of the absolute value between the pre-ranking and ranking model, however, it does not assess the ranking quality of the model.
In practice, a ranking model, e.g., CTR model, is commonly assessed with the ranking quality on the win set. In terms of the pre-ranking CTR model, 
we find that aligning the ranking quality of pre-ranking model with the ranking model also helps increase the ranking quality.
One way to measure the ranking quality of the single-objective pre-ranking model is to use the RCS with the single objective and set the fusion function $g$ as the identity function, i.e., $G(X) = g(F_i(X)) = F_i(X)$, where the $i$-th objetive model is evaluated. In this scenario, the single-objective RCS measure the alignment of the ranking quality between the pre-ranking and ranking model. 

\subsection{Revealing the Cause of the Inconsistency of Pre-ranking Stage}

Given the proposed RCS, we can also use it  to find the module that causes the inconsistency. Concretely, we can calculate RCS between the ranking stage and pre-ranking stage. Ideally, the RCS should be close to 1.0, otherwise, the inconsistency exists. Then we can substitute each model with the corresponding model from the ranking stage, and compute the new RCS. 
If replacing one model can raise RCS drastically, then the pre-ranking model is the cause of the inconsistency (RCS is 1.0 if replacing all models). For example, in Table~\ref{tab:rcexample} if we replace the pre-pCTR model with the pCTR model, the RCS increase to 1.0 which suggest the pre-pCTR model is the cause of inconsistency. 

After finding the model that causes the inconsistency, we can improves the model by increasing the ranking quality and the proxy-calibration. In section~\ref{sec:methodology}, we propose several methods for improving the ranking consistency.

\begin{figure*}[!t]
    \centering
    \includegraphics[width=0.85\textwidth]{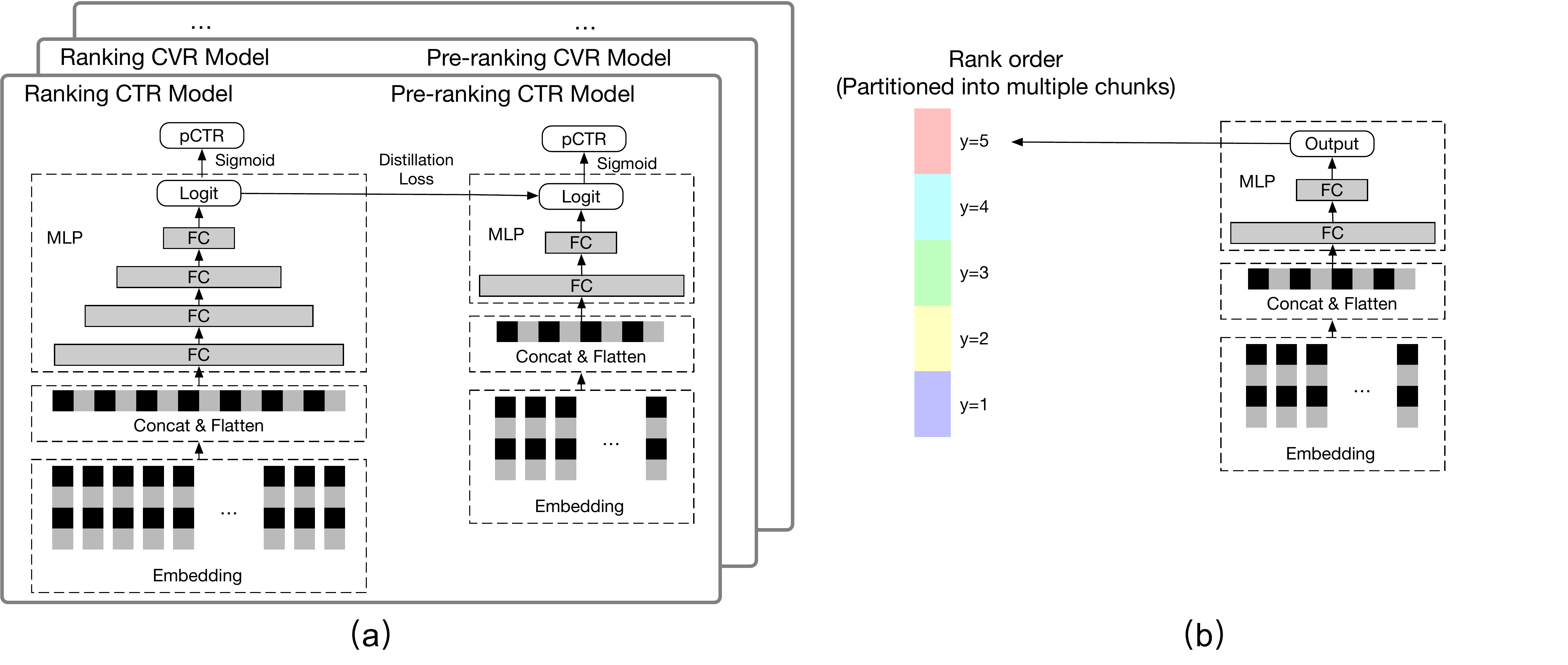}
    \caption{(a) Improving proxy-calibration by distilling multiple models. (b) One model to learn the rank order.}
    \label{fig:method}
\end{figure*}

\section{Methodology}\label{sec:methodology}
 
In this section, we aim to improve the ranking consistency from the perspective of sample selection and learning algorithm.

\subsection{Sample Selection}

In industrial ranking system, exposed samples in the win set are often collected to train models. However, when serving online, the pre-ranking model needs to give prediction scores for items in the pre-ranking set (most of which are unexposed). The mismatch between training and serving samples is known as the Sample Selection Bias (SSB)~\cite{MaZHWHZG2018ESMM}. We discover that, the SSB problem can also leads to ranking inconsistency. In detail, although the distribution of prediction scores on the win set between the pre-ranking and ranking stages are similar, it is quite different on the pre-ranking set. We will discuss the result in Section~\ref{sec:ranking_inconsistency}. The result suggests that the inconsistency are more severe on unexposed samples since they are not included in the training set.

To mitigate the inconsistency caused by SSB, we propose to utilize samples in the \textbf{pre-ranking set} to train the pre-ranking models. However, there are two difficulties: 1) we do not have user engagements as labels on unexposed samples. 2) The pre-ranking set is hundreds (or even thousands) times as large as the win set. It is hard for the model to train all samples in real-time. 
For the first difficulty, recall that the goal of the ranking consistency is to align the pre-ranking stage with the ranking stage, so we can use the prediction scores of ranking models as the proxy labels. Concretely,
for each user requests, the ranking models deployed on the online simulator give the prediction score for each item in the pre-ranking set, which will be used as the label to train the pre-ranking models.
For the second difficulty, we can use various downsampling strategy to reduce the amount of training samples. For example, one can uniformly sampling from the pre-ranking set or uniformly sampling from the smaller competitive set to construct the training set.

\subsection{Learning Algorithms}
In this section, we propose to two learning algorithms for improving the ranking consistency: 1) distilling the prediction scores of ranking models for improving the proxy-calibration, 2) directly learning the rank order of the ranking stage.

\subsubsection{Improving Proxy-calibration By Distilling Multiple Models}
One way to improve the ranking consistency is to improve the proxy-calibration of each pre-ranking model, i.e., let the prediction of each single pre-ranking model as close as that of the corresponding ranking model. 
Concretely, we employ the downsampled pre-ranking set as the training set with the prediction of the ranking model as the proxy label. The pre-ranking model is trained to fit the proxy label.
Note that this method can be seen as a specific knowledge distillation method~\cite{2015Distilling} that transfer the knowledge from
the teacher model (ranking model) to the student model (pre-ranking model) on the pre-ranking set.  Unlike previous work on distilling knowledge in recommendation models that focus on distill knowledge from the win set ~\cite{XuLGGYPSWSO2020PFD,DBLP:conf/kdd/TangW2018RankingDistillation}, the main difference of the proposed model is that we consider the pre-ranking set and aim to improve the proxy-calibration for ranking consistency. 
The model architecture is illustrated in Figure~\ref{fig:method}(a). 
The logit~\footnote{Logit is referred to as the model output before the last activation layer (e.g.,
sigmoid or softmax function).} of ranking model are used as the proxy label and mean square error are employed as the loss function:
\begin{equation}
\begin{aligned} 
    \frac{1}{N}\sum_i (\rm{logit_{ranking}} - \rm{logit_{pre-ranking}})^2,\\ 
\label{eq:distill}
\end{aligned}
\end{equation}
where $N$ is the number of samples. 
The training objective in Equation~\ref{eq:distill} is to fit the logit of the ranking model. During serving, an activation layer (e.g., sigmoid for CTR model) is added to get the final output.

The proposed method is an end-to-end calibration model for the pre-ranking stage, the target of which is to keep the consistency between the pre-ranking and ranking stage. Note that other distillation objective or post calibration methods, such as isotonic regression~\cite{ZadroznyE2002Isotonic}, can also be explored to improve the proxy-calibration, which is open for further research. 

\subsubsection{One Model to Learn the Rank Order}
The common design of pre-ranking stage is to build separate models for different objectives like the ranking stage. In this manner, the rank order of pre-ranking stage is more likely to be inconsistent since the error of different models are also fused and amplified. Note that the goal of the pre-ranking stage is to select as much as items in the ideal win set for the ranking stage. 
Thus, to improve the ranking consistency, we can reduce the task complexity and use one model to directly learn the order of ranking stage. 
Concretely, we use the downsampled pre-ranking set as the training set and compute the fused ranking score using the predictions of ranking models. Then for each request, we rank the training data by the fused ranking score and cast the problem as a learning to rank (LTR) problem. 
The goal of the pre-ranking model is learning to rank the pre-ranking set as the ranking stage.
To further reduce the learning cost, the list ranked by the fused rank score is partitioned into several chunks, each containing a group of items with adjacent rank positions. For example, we can treat top items that are exposed as the positives $y=1$ and the others as negatives $y=0$. A more general way is to use $k$ chunks, $y\in \{1, \dots, k\}$ and learn to discriminate items among chunks.
For the loss function, we use the pairwise RankNet loss~\cite{2010From} as Equation~\ref{eq:ranknet}:
\begin{equation}
\begin{aligned}
-\sum_{i\neq j}\mathbb{I}_{y_i>y_j} \log\frac{\exp{(s_i-s_j)}}{1+\exp{(s_i-s_j)}},
\label{eq:ranknet}
\end{aligned}
\end{equation}
where $y_i, y_j$ is the trunk indicator of the $i$-th and $j$-th sample. The architecture is shown in Figure~\ref{fig:method}(b). 
By doing so, we use one model to fit the rank order directly instead of fitting each objective separately.

\section{Experiments}
In this section,  we conduct experiments in the real world display advertising system to validate the following arguments:
\begin{itemize}
    \item The ranking consistency affects online performance.
    \item The proxy-calibration and ranking quality of single-objective model are important factors that affect the ranking consistency.
    \item By using the proposed RCS as metric, we can find the inconsistency module of pre-ranking stage. We further show that the proposed methods can improve the ranking consistency and online performance.
\end{itemize}
Note that although it is a specific advertising system, the analysis methodology \textbf{generalizes} to any stage of any ranking system.

\subsection{Setup}
In the advertising system, the final objective is the effective Cost Per Mille (eCPM) that is the product of bid and pCTR. Here, the bid and pCTR are two single objectives. The bid can be the init-bid or the opt-bid. The difference is listed as follow
\begin{itemize}
\item \textbf{init-bid} is the initial bid provided by advertiser for an ad.
\item \textbf{opt-bid} is the optimized bid based on the initial bid by our system. In our advertising system, we may help advertiser raise or reduce the bid according to the quality of current traffic~\cite{ZhuJTPZLG2017OCPC}. Specifically, {opt-bid} is a function of init-bid, pCVR, and other information such as budgets of the advertiser.
\end{itemize}
In the ranking stage, the pCTR is estimated by a complicated model, which we will refer to as \textit{Rank} model. For the pre-ranking stage, we will compare five models,  \textit{Logloss}, \textit{Logloss-small}, \textit{Logloss-med}, \textit{Distill-Pre} and \textit{LTR}. For simplicity, the output of these models will be indicated as |Rank|, |Logloss|, |Logloss-small|, |Logloss-med|, |Distill-Pre| and |LTR|, respectively.
The details of these models are listed as follow
\begin{itemize}
\item \textbf{Rank} is the pCTR model at the ranking stage. We use CAN~\cite{Zhou2020CAN} as the model architecture. The training set is the win set. For each sample, the label is 1 if the user click the ad and 0 otherwise.
\item \textbf{Logloss} is a condense version of the Rank model. The only difference is that it use less features and a simpler architecture. 
The Logloss model is also trained on the win set and use the logloss as the loss function.
\item \textbf{Logloss-small} has a much smaller set of feature compared with  Logloss model, the model architecture and training manner is as same as the Logloss model.
\item \textbf{Logloss-med} has a medium set of feature, which is larger than the feature set of Logloss-small but smaller than the feature set of Logloss.
The model architecture and training manner is as same as the Logloss model.
\item \textbf{Distill-Pre} is the proposed distillation model trained on the downsampled pre-ranking set. In the experiment, instead of directly downsampling from the pre-ranking set, we take the competitive set as the training set.
\item \textbf{LTR} is the proposed model that learns the rank order. The competitive set is used as the training set. Note that 
in advertising system, we should consider the Incentive Compatibility (IC) and Individual Rationality (IR). To satisfy these properties, the system need to ensure the winning chance of an ad will not be lower if the advertiser raise the bid~\cite{wilkens2017gsp}. To ensure that, we decompose the rank score at pre-rank stage as init-bid*|LTR|, where the objective of |LTR| is fitting the $\frac{\textrm{|Rank|*opt-bid}}{\textrm{init-bid}}$. During inference, the bid provided by the advertiser will be used for the pre-ranking stage. 
\end{itemize}

\textbf{Implementation Details} All models are trained using data from the past 60 days. The Logloss, Distill-Pre and the LTR share the same feature set and the same model architecture, and hence are fair on capacity.



\subsection{Ranking Consistency}
We deploy all these models on the display advertising system for A/B testing. The RCS are computed using online service and simulator logs of three days.  
In our experiment, we use the opt-bid and the Rank model for the ranking stage
and varying the pre-ranking models to observe the relation between ranking consistency and online performance. For the online metric, we calculate the relative improvement on CTR and RPM of each model against the Logloss model, which is the production baseline.
The result is shown in Table~\ref{tab:exp_setsrecall}.
Our first observation is that \textbf{the online performance is consistent with the ranking consistency.} 
Compared with the logloss model, Logloss-small and Logloss-med have inferior RCS and worse online performance due to the use of less features. Additionaly, LTR and Distill-Pre method are more consistent with the ranking stage and improve the RCS from 64.1\% to 77.2\% and 79.7\%, respectively.
During the online performance, we observe 5.1\% and 5.5\% improvement on RPM of these two methods. This verifies the first argument that ranking consistency affects online performance. 
The performance gain also demonstrates that \textbf{Distill-Pre and LTR effectively improve the ranking consistency}, which validates the third argument.

\begin{table}[!tbp]
  \caption{Performance comparisons to verify the alignment between ranking consistency and online performance.}
  \label{tab:exp_setsrecall}
  \resizebox{\columnwidth}{!}{
  {\begin{tabular}{c|c|c|c|c}
  
  \toprule
 
    \multicolumn{2}{c|}{Ranking Rule} & \multicolumn{3}{c}{Performance} \cr
 
    \midrule

  Pre-ranking stage&  Ranking stage& RCS  & CTR & RPM \\
  \midrule
  init-bid* |Logloss-small| & opt-bid*|Rank|       & 60.4\%               & -2.0\%         & -0.7\%           \\
  init-bid* |Logloss-med| & opt-bid*|Rank|       & 63.1\%               & -0.2\%         & -0.1\%           \\
init-bid* |Logloss| & opt-bid*|Rank|       & 64.1\%               & +0.0\%         & +0.0\%           \\
init-bid*|LTR| & opt-bid*|Rank| & 77.2\%            & +6.2\%         & +5.1\%            \\
init-bid*|Distill-Pre| & opt-bid*|Rank| & 79.7\%             & +6.7\%         & +5.5\%           \\

  \bottomrule
  \end{tabular}}
  }
\end{table}

\subsection{Using RCS to Find Inconsistency Models}
We show that the proposed RCS metric can help find the model
that causes inconsistency.  Here, the baseline uses init-bid and Logloss model for the pre-ranking stage. Using opt-bid and Rank model in the pre-ranking stage can be seen as the performance upper bound, in which case the RCS is 100\% and the system is perfectly consistent.
We conduct two experiment by replacing each pre-ranking model with corresponding ranking model, i.e., 1) replace the Logloss with the Rank model and 2) replace the init-bid with the opt-bid. Then we calculate the new RCS after the replacement.
The result is shown in Table~\ref{tab:exp_local}. 
Our first observation is that both the Logloss model and init-bid can lead to the inconsistency. By replacing the Logloss model with the Rank model, the RCS increases from 64.1\% to 95.6\%. The RCS also improves from 64.1\% to 85.9\% if we replace the init-bid with the opt-bid. 
The result also suggest that it is the baseline CTR model Logloss that mainly causes the inconsistency in current system. By optimizing the CTR model, we can significantly improving the consistency of the system. 

\begin{table}[!tbp]
  \caption{Performance comparisons to reveal ranking inconsistency.}
  \label{tab:exp_local}

  {\begin{tabular}{c|c|c}
  
  \toprule

    \multicolumn{2}{c|}{Ranking rule} & \multicolumn{1}{c}{Performance} \cr
    \midrule

  Pre-ranking stage&  Ranking stage& RCS    \\
  \midrule
init-bid*|Logloss| & opt-bid*|Rank| & 64.1\%                              \\
\midrule
init-bid*|Rank| & opt-bid*|Rank| & 95.6 \%                \\
opt-bid*|Logloss| & opt-bid*|Rank| & 85.9 \%                \\
\midrule
opt-bid*|Rank| & opt-bid*|Rank| & 100.0 \%                \\

  \bottomrule
  \end{tabular}}
\end{table}

 \begin{table}[!tbp]
  \caption{Performance comparisons to evaluate the quality of ranking.}
  \label{tab:exp_quality_ranking}
  {\begin{tabular}{c|c|c}
  
  \toprule

   Pre-ranking model & Single-Objective RCS & RCS  \\
  \midrule
|Logloss-small|    &75.4\% & 60.4 \%                              \\
|Logloss-med|    &78.7\%  & 63.1\%                         \\
|Logloss|  & 87.3\% & 64.1\%                   \\
|LTR| & 85.4\% & 77.2\%\\ 
|Distill-Pre| & 95.3\% & 79.7\%\\ 
  \bottomrule
  \end{tabular}}
 \end{table}
\subsection{Factors that Influence the Consistency}~\label{sec:ranking_inconsistency}
We investigate the factors that relates to the ranking consistency when iterating single-objective model. 

\begin{figure*}[!t]
    \centering
    \begin{minipage}[h]{.4\textwidth}
        \centering
        \includegraphics[width=\columnwidth]{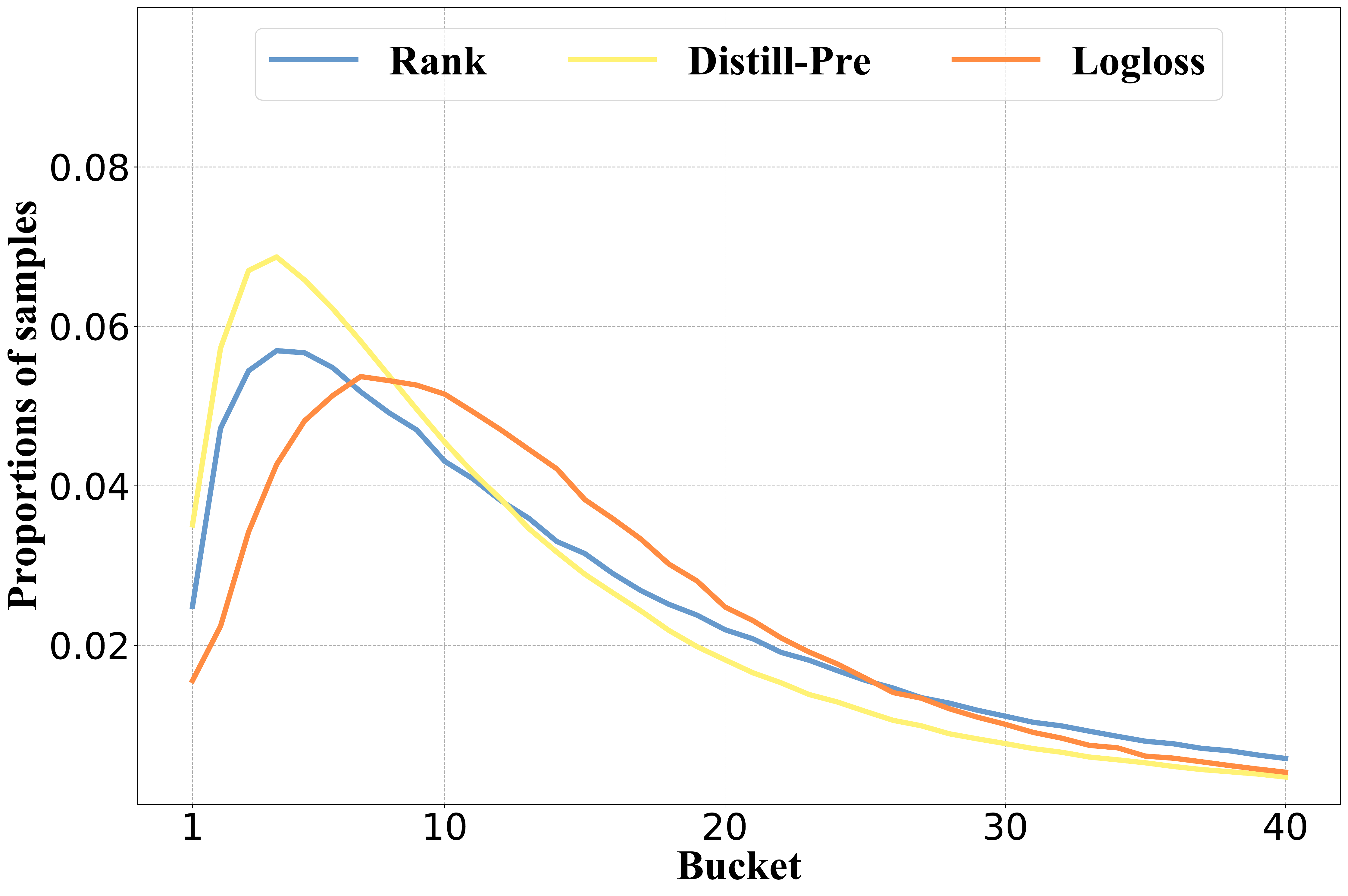}

        \mbox{({\it a}) {Win set}}
    \end{minipage}
    \begin{minipage}[h]{.4\textwidth}
        \centering
\includegraphics[width=\columnwidth]{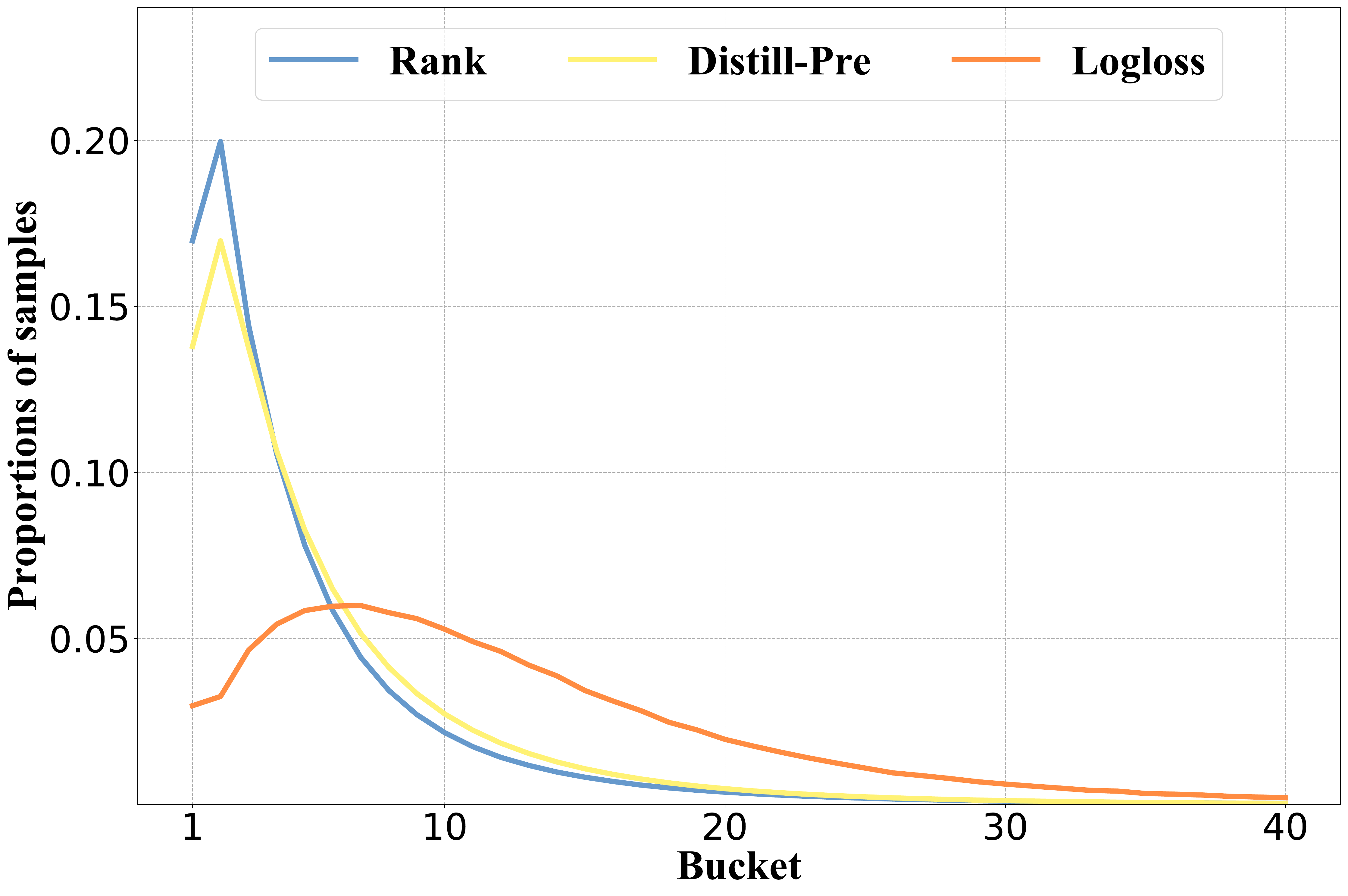}
        \mbox{ ({\it b}) {Pre-ranking set.}}
    \end{minipage}
    \caption{The distribution plots of prediction scores on Win set and Pre-ranking set.}
    \label{fig:pdf}
\end{figure*}

\subsubsection{Ranking Quality}
We first study the relation between ranking quality of single-objective model and the ranking consistency. 
Here we measure the ranking quality of each pre-ranking pCTR model using the single-objective RCS metric.
Different with Equation~\ref{eq:RCS}, the single-objective RCS use the pCTR of the pre-ranking set for ranking but not the eCPM (pCTR*bid) for ranking. The single-objective RCS measure the alignment of the ranking quality between the pre-ranking pCTR model and ranking pCTR model.

The result is illustrated in Table~\ref{tab:exp_quality_ranking}. 
We observe that the ranking quality of pCTR model is closely related to the ranking consistency. For the Logloss-small, Logloss-med, Logloss and the Distill-Pre model, the RCS increases with the increase of ranking quality (single-objective RCS). One exception is the LTR model - compared with the Logloss model, the single-objective RCS of LTR drop from 87.3\% to 85.4\%, but the RCS increase from 64.1\% to 77.2\%. The reason is that the LTR model is not actually a pCTR model since it is fitting  $\frac{\textrm{|Rank|*opt-bid}}{\textrm{init-bid}}$. Consequently, the ranking quality of the LTR model is not consistent with the pCTR model.

  


\subsubsection{Proxy-Calibration}
Besides the ranking quality, the proxy-calibration also relates to the ranking consistency. The reason is that the scale of the prediction scores matters when fusing multiple objectives. 
To measure the proxy-calibration performance, we use the Expected Calibration Error (ECE) as the metric.  For the $i$-th sample, let $p_i$ and $\hat{p}_i$ represents the predicted pCTR of the ranking model  and pre-ranking model, respectively.
To compute ECE, we first partition the range [0, 1) equally into $K$ buckets (K=50). $\vmathbb{1}(\hat{p}_i\in{B_k})$ is an indicator with a value of 1 if the pCTR locates in the $k$-th bucket $B_k$, and otherwise 0. ECE is computed as  Equation~\ref{eq:ECE}: 
\begin{equation}
    \text{ECE} = \frac{1}{D}\sum_{k=1}^{K}|\sum_{i=1}^{D}(p_i-\hat{p}_i)\:\vmathbb{1}(\hat{p}_i\in{B_k})|,
    \label{eq:ECE}
\end{equation}
where $D$ is the number of samples. 
Note that a lower ECE implies a better performance.
The ECE of the Logloss and the Distill-Pre model are listed in Table~\ref{tab:ece}. The result shows that the Distill-Pre model significantly outperforms the Logloss model on proxy-calibration.

\begin{table}[!tbp]
\centering
\caption{ECE of the  Logloss and Distil-Pre.}
\label{tab:ece}
{
\begin{tabular}{c|c|c}
    \toprule
            &   Logloss & Distill-Pre\\
    \midrule
    ECE    & 0.3070 & 0.0318\\
    \bottomrule
\end{tabular}
}
\end{table}

To understand the cause of the poor proxy-calibration performance of Logloss model,
we visualize the pCTR distribution of Rank, Distill-Pre and Logloss model on the win set and pre-ranking set. 
Concretely, we partition the range [0, 1) equally into K buckets, and each sample is located in a bucket according to its predicted probability. Then we compute the proportion of samples of each bucket, as shown in Figure~\ref{fig:pdf}. We can see the plots of distribution density of Rank and Logloss model on competitive set differ from each other substantially, while the plots on win set look much more similar. The results imply that the poor proxy-calibration of the Logloss model are mainly caused by unexposed samples since they are not included in the training set. In contrast, the Distill-Pre model use the pre-ranking set as training set and the pCTR of Rank model as the proxy-label for training, resulting in better proxy-calibration on  pre-ranking set.

  


\section{Conclusion and Future Work}
In this paper, we study the problem of ranking consistency between the pre-ranking and ranking stages and propose the Ranking Consistency Score (RCS) as the evaluation metric. We find that the ranking consistency has a strong impact on the performance of industrial systems. 
We also discover that both
the ranking quality and proxy-calibration of single objective
influence the ranking consistency.
To improve the consistency, we propose sample selection and learning algorithms. The experiment result validates the efficacy of the proposed methods.
Up to now, the proposed method has been examined through A/B test on the display advertising system, obtaining a 6.7\% improvement on CTR and a 5.5\% increase on RPM.

The results also suggest several future research directions. Firstly, ranking consistency should be considered in all stages, e.g., retrieval stage. We can also define the RCS metric between the retrieval and pre-ranking stages. Ideally, if every two adjacent stages are consistent, then the consistency of the whole ranking system is achieved. Secondly, although consistency of all stages can enhance the utility, it may also make the ranking system biased toward items that are frequently exposed. To alleviate this issue, explore strategy should also be investigated to ensure the long-term profit.


\bibliographystyle{ACM-Reference-Format}
\balance
\bibliography{rn_odl}


\end{document}